%% file: apssamp_letter_arxiv.tex
\documentclass[%
 reprint,
superscriptaddress,
nofootinbib,
 amsmath,amssymb,
 aps,
]{revtex4-2}

\usepackage[T1]{fontenc}
\usepackage{lmodern}
\usepackage[utf8]{inputenc}
\usepackage[table,dvipsnames]{xcolor}
\usepackage{multirow}
\usepackage{graphicx}
\usepackage{mathtools}
\usepackage{amsmath,amssymb,amsfonts,amsthm,latexsym,stmaryrd,physics,mathrsfs}
\allowdisplaybreaks[4]
\usepackage{marginnote}
\usepackage{color}
\usepackage{bm}
\usepackage{longtable}
\usepackage{tikz}
\usepackage{float}
\usepackage{nicefrac}
\usepackage{microtype} 
\usepackage{booktabs}
\usepackage[ruled,linesnumbered]{algorithm2e}

\usepackage{verbatim}
\usepackage{setspace}
\usepackage[T1]{fontenc}
\usepackage[utf8]{inputenc}
\usepackage{stfloats}
\usepackage{diagbox}
\usepackage{dsfont}
\usepackage{hyperref}
\usepackage[colorinlistoftodos,prependcaption,textsize=tiny]{todonotes}
\usepackage{epstopdf}
\usepackage{latexsym}
\usepackage{makecell}
\usepackage{xparse}
\usepackage[center]{subfigure}

\usetikzlibrary{positioning, shapes.geometric}

\usepackage{xargs}                      %
\usepackage[colorinlistoftodos,prependcaption,textsize=tiny]{todonotes}
	
\usepackage{ulem} %

\input{defs.tex}

\graphicspath{{./figs}}

\newcommand{\RN}[1]{%
\textup{\uppercase\expandafter{\romannumeral#1}}%
}

\begin{document}
\title{From Principles to Effective Models: A Constructive Framework for
Effective Covariant Actions with a Unique Vacuum Solution
}%

\author{Kristina Giesel}
\email{kristina.giesel@fau.de} 
\affiliation{Department of Physics, Institute for Quantum Gravity, Theoretical Physics III, Friedrich-Alexander Universit\"at Erlangen-N\"urnberg, Staudtstr. 7, 91058 Erlangen, Germany}

\author{Hongguang Liu}
\email{liuhongguang@westlake.edu.cn}
\thanks{Corresponding author}
\affiliation{Institute for Theoretical Sciences and Department of Physics, Westlake University, Hangzhou 310024, Zhejiang, China}
\affiliation{Department of Physics, Institute for Quantum Gravity, Theoretical Physics III, Friedrich-Alexander Universit\"at Erlangen-N\"urnberg, Staudtstr. 7, 91058 Erlangen, Germany}

\date{\today}

\begin{abstract}
The absence of Birkhoff's theorem in effective quantum gravity models  leads to a fundamental ambiguity in the vacuum sector, where a priori no unique vacuum solution exists. As a result, phenomenological investigations of the physical implications of these models have been made more difficult. We address this challenge by establishing a constructive framework which allows to formulate $4$D covariant actions from the physical nature of the systems's degrees of freedom, which are dust and gravity, together with two guiding principles. We take advantage of the non-propagating nature of a relational dust clock and the suppression of gravitational waves in spherical symmetry. This structural ultralocality allows for a decomposition of the dynamics into independent LTB shells. We further impose spatial diffeomorphism invariance and a geometric guiding principle, where the latter ensures that a unique and static vacuum solution exists. These assumptions allow to strictly constrain the LTB shell Hamiltonian to a factorised form as well as the static vacuum metric function to a universal form. This constructive framework produces a fully 4D-covariant action that belongs to the class of generalised extended mimetic gravity models. This provides the necessary consistent basis for a perturbation theory in the context of quasi-normal modes or cosmological perturbations beyond the static sector in which quantum gravity effects are also included in linear and higher order perturbations. Furthermore, for this class of models our results resolve the long-standing ‘curvature polymerisation ambiguity’ in loop quantum cosmology by unambiguously determining how flat space modifications are extended to non-flat geometries, thus unifying the description of black holes and cosmology in a single effective framework.
\end{abstract}

\maketitle

\section*{Introduction}\label{sec:intro}

Birkhoff's  theorem  guarantees that every vacuum solution of Einstein's equations with spherical symmetry must be static and asymptotically flat and is therefore,up to coordinate transformations, uniquely given by the Schwarzschild solution,  determined solely by a constant Misner-Sharp mass $M$. When dust is added and one chooses the comoving gauge, the work in \cite{Lasky:2006hq} shows that all spherically symmetric solutions with dust correspond to the LTB solution up to coordinate transformations. LTB solutions are characterised by two free functions $\mathcal{E}(x)$ and $M(x)$ encoding the energy/spatial curvatutre profile and the Misner-Sharp mass inside a dust shell respectively, where $x$ denotes the radial coordinate in LTB coordinates and $\mathcal{E}(x)$  is often denoted as the LTB function. For marginally bound LTB solutions ($\mathcal{E}(x)=0$) Schwarzschild solutions can be easily embedded into LTB solutions by choosing $M(x)=M$ and Birkhoff's theorem can be directly rediscovered. 
This is usually how polymerised vacuum solutions or the Birkhoff theorem is discussed in effective models, e.g. \cite{Kelly:2020uwj,Giesel:2024mps}. In the non-marginally bound case ($\mathcal{E}(x)\not=0$) from the LTB perspective one obtains a family of vacuum solutions characterised by $\mathcal{E}(x)$. In classical GR the presence of $\mathcal{E}(x)$ is usually interpreted as pure gauge since neither the coordinate invariant form of the Misner-Sharp mass nor the Kretschmar or  Weyl scalar depend on it and a time-like Killing vector exists also if $\mathcal{E}(x)\not=0$. As discussed in \cite{Lasky:2006hq} different choices for $\mathcal{E}(x)$ correspond to changes in  the relationship between the comoving proper time and Schwarzschild Killing time, i.e. the foliation. Hence, while the LTB solution in GR also only encodes a unique physical solution, the Schwarzschild one, the form of the metric encodes a family of slicings related to different free-falling radial observers. The family of vacuum solutions in the LTB context is structural in 4 dimensions when Lovelock's theorem confirms the uniqueness of GR up to topological terms \cite{Lovelock:1971yv,Clifton:2011jh}; in contrast, lower-dimensional (symmetry-reduced) models, e.g. 2d dilaton-gravity models, often exhibit automatic Birkhoff-type uniqueness due to their topological nature \cite{Grumiller:2002nm,Witten:2020ert}, for which a detailed analysis in the Hamiltonian framework can be found in \cite{Zhang:2025ccx}.

However, the situation changes drastically when we move to effective models of quantum gravity, such as those inspired by Loop Quantum Gravity (LQG) or bouncing cosmology.
This ambiguity plays a different role than in classical GR \cite{Ashtekar:2023cod,Giesel:2023hys,Giesel:2023tsj,Giesel:2024mps,Cafaro:2024vrw}. %
As the explicit calculation of the curvature scalar and the Weyl scalar shows, these quantities do indeed depend on the LTB function $\mathcal{E}$ in the class of effective models under consideration, e.g. \cite{Cafaro:2024vrw}. Consequently, the presence of $\mathcal{E}$ in the polymerised vacuum solution cannot simply be interpreted as a pure gauge artifact, as in classical GR. %
This leads to a profound `uniqueness problem': without further guiding principles, the effective dynamics admit an infinite family of static vacuum solutions parameterisd by $\mathcal{E}(x)$. As a consequence, phenomenological investigation for such models depend on the specific choice of this parametrisation. However, given the restrictions imposed by Lovelock's theorem, it is not trivial to solve this purely at the level of a covariant 4D action formulation.

In this work, %
our aim is not just to find a solution, but to {rediscover the uniqueness} of the vacuum solution in effective gravity and, crucially, to use this uniqueness to \emph{construct the underlying covariant theory}. We generalise the reconstruction scheme of \cite{Giesel:2024mps} to the non-marginally bound case ($\mathcal{E} \neq 0$) and demand that the physical requirements of the theory must restore a generalised Birkhoff theorem.

Our framework is characterised by its physical assumptions. Instead of imposing a mathematical ansatz for a spherically symmetric Hamiltonian, we derive the action of the model from two fundamental physical requirements regarding the degrees of freedom in the effective model:
\begin{itemize}
    \item {Matter Sector:} We maintain the dust description for the matter content. Due to its pressureless nature, the dust (as an additional scalar degree of freedom) fluid possesses a vanishing sound speed ($c_s = 0$). This implies no propagation of mechanical disturbances between fluid layers. It has a clear effective vacuum limit, where the dust density tends to zero but still we keep the effective dynamics, which we denote as polymerised vacuum solution. In such a case, the dust behaves as test particles following geodesics. Moreover, likewise to 
    \cite{Kuchar:1990vy,Kiefer:2005tw} in the case of a non-vanishing dust density we can choose dust as a relational clock field, which allows a straightforward construction of a physical Hamiltonian \cite{Giesel:2007wn,Ferrero:2025est}.%

    \item {Gravity Sector:} We assume that the modification to the gravitational Hamiltonian is such that it introduces no new propagating degrees of freedom. This ensures that the total count of degrees of freedom in the full 4D covariant theory remains $2 \text{ (tensor)} + 1 \text{ (scalar/dust)}$. Since the scalar mode is non-propagating ($c_s=0$) and the tensor modes (gravitational waves) are  suppressed in spherical symmetry, the system exhibits \emph{structural ultralocality} as independent shell Hamiltonian when we reduce to spherical symmetry.
\end{itemize}

To ensure the uniqueness of the vacuum solution for a given effective model, we extend our constructive framework by further physical requirements: (i) \emph{3D spatial diffeomorphism invariance}: This requirement enforces the momentum constraints, ensuring a strict $2+1$ degree of freedom count (tensor + scalar) and enabling the consistent construction of the effective dynamics into a full {4D covariant action}. (ii) \emph{Geometric guiding principle}: Despite the enlarged phase space from the relational clock field, we want the polymerised vacuum solution to be static and unique, mimicking Birkhoff's theorem of its classical counter part. As a consequence, we require that it can only depend on one parameter, the constant Misner-Sharp mass.

We show that for models with dust, the resulting four-dimensional covariant theory belongs to a class of \emph{generalised extended mimetic gravity models}, which extends the previous extended mimetic gravity models to include corrections depending on three curvature scalars \cite{Takahashi:2017pje, Langlois:2018jdg}. 

It is worth noting that this result can be further generalised to theories with a scalar field clock encoding non-vanishing pressure, as will be detailed in a forthcoming work \cite{Liu:2026scalar}. However, in such cases, the notion of a unique `polymerised vacuum' becomes ill-defined, as the scalar field with pressure introduces extra propagating degrees of freedom, breaking the ultralocality essential for the Birkhoff-type uniqueness found in the dust case. %

With the decomposition into independent shells established by the structural ultralocality, we then apply our physical principles to constrain the explicit form of the dynamics. First, the requirement of \emph{spatial diffeomorphism invariance}  restricts the shell Hamiltonian to a specific factorised form given by $H[b,v,\mathcal{E}] = v H_0(b, \mathcal{R}_1={\mathcal{E}}/{v^{2/3}})$, where $b$ is the connection variable and $v$ the conjugate volume variable used in LQC models in cosmology. Second, the \emph{geometric guiding principle} ensures that none of the curvature scalars nor the Weyl scalar depends on the LTB function $\mathcal{E}$. In our framework, this requirement translates into a condition on the shift vector in PG coordinates. Combining the latter with the effective dynamics 
leads to a partial differential equation linking $H$ to the static metric function $f(M,r)$, where $(\tau,r)$ denotes Schwarzschild-like coordinates that exactly agree with Schwarzschild coordinates for $f(M,r)=\frac{2M}{r}$, where we set Newton's constant $G=1$ everywhere. Combining then (i) and (ii), we derive the \emph{universal functional form} for any static polymerised vacuum solution admitted by such a model:
\begin{align}\label{eq:main_result_boxed}
\boxed{\; f(M, r) = r^{2} \; \widetilde{f}\!\left( \frac{2M}{r^{3}} \right) \;},
\end{align}
where we set Newton's constant $G=1$ everywhere. The Schwarzschild metric, $f=2M/r$, corresponds to the linear choice $\widetilde{f}(X)=X$. \eqref{eq:main_result_boxed} states that modifications can only enter as a function of the dimensionless mass-density scale $M/r^{3}$, multiplied by the geometric factor $r^{2}$. Different choices of $\widetilde{f}$ correspond to different models each equipped with a unique vacuum solution. Many existing effective black hole solutions can be embedded in this universal form, including Hayward \cite{Hayward:2005gi}, Bardeen \cite{Bardeen68}, Asymptotically safe gravity \cite{Pawlowski:2023dda} and LQG inspired models \cite{Lewandowski:2022zce,Zhang:2025ccx}.

This result has two far-reaching implications. First, it provides a unique static vacuum solution completely determined by its mass at the effective level.  Then through the derived spatially covariant action, that can be extended to a full 4D generalised extended mimetic  theory, it enables a consistent perturbation theory analysis  beyond the hybrid approximation \cite{ElizagaNavascues:2020uyf}, where only the background solutions is modified by polymerisations. Second, and crucially for LQC, in the case of an (effective) dust clock it resolves the persistent ‘curvature polymerisation ambiguity’: the form $H = v H_0(b, \mathcal{R}_1)$ uniquely determines how the modified dynamics of flat space cosmology must be extended to account for spatial curvature in case we want to construct the effective spherically symmetric model with a unique vacuum solution from infinitely many decoupled non-flat FRW models. Our construction implies a unified description of effective gravity. The same covariant action that describes the polymerised black hole interior and exterior also governs the evolution of cosmological models. This allows for a seamless transition between black hole physics and cosmology within a single, consistent effective framework.

After a minimal review of spherically symmetric dust models based on decoupled LTB models, we %
construct the framework using our physical principles and
show how the constraint \eqref{eq:main_result_boxed} can be derived, discuss the resulting covariant theory, and illustrate its capability with an explicit example from polymerised LQG-inspired models.

\section*{Reconstruction Algorithm: Extended to the Non-Marginally Bound case}
The class of effective spherically symmetric models is based upon LTB dust models that have decoupled shell dynamics investigated in \cite{Giesel:2023tsj}. In comoving coordinates, each dust shell is described by a canonical pair in connection variables $(b, v)$ with canonical Poisson brackets satisfying $\{b,v \} =3/2$, where $v \equiv R^{3}$ and $R(t,x)$ is the areal radius and we denote the radial coordinate by $x$ and comoving time by $t$, together with a conserved LTB function $\mathcal{E}(x)$ \cite{Szekeres:1999,Lasky:2006hq}. The shell Hamiltonian $H[b, v, \mathcal{E}]$ generates the evolution \cite{Kiefer:2005tw, Bojowald:2009ih,Giesel:2021dug,Giesel:2024mps}. In GR, it takes the form $H_{\text{GR}} = b^{2} v - \mathcal{E} v^{1/3}$.

The full spherically symmetric geometry can recovered through a constructive framework, as a generalization of the reconstruction procedure for marginally bound case ($\mathcal{E}=0$) introduced in \cite{Giesel:2023hys,Giesel:2024mps} %
, which expresses $\mathcal{E}$ and $H$---and via the equations of motion, $b$---in terms of the standard connection canonical variables of the spatial metric \cite{Bengtsson_1990,Bojowald:1999eh,Bojowald:2004af}, via the following identification map $\mathcal{I}$:
\begin{align}
\mathcal{I}(\mathcal{E}) &\quad \longrightarrow\quad \mathcal{I}(\mathcal{E})[E] = \left( \frac{(E^x)'}{2 E^\phi} \right)^2 - 1, \label{eq:map_E_to_full} \\
\mathcal{I}(v = R^3)&\quad \longrightarrow\quad (E^x)^{3/2}, \\
\mathcal{I}(b) &\quad \longrightarrow\quad  \frac{K_{\phi}}{\sqrt{E^x}}, \label{eq:map_Kphi_to_full} \\
\mathcal{I}(b') &\quad \longrightarrow\quad \frac{(E^x)' (2 \sgn((E^x)') E^x K_x - E^{\phi} (K_{\phi})}{2 E^{\phi} (E^x)^{3/2}} \label{eq:map_Kx_to_full},
\end{align}
where the prime denotes the radial derivative and the phase space involves extrinsic curvature components $K_\phi,K_x$ as well as their momenta the densitised triads $E^\phi,E^x$. The connection variables satisfy the canonical Poisson algebra $\{K_\phi(x), E^{\phi}(y)\} = \{K_x(x), E^{x}(y)\} = \delta(x-y)$ \cite{Giesel:2023tsj} where all other Poisson brackets vanish.
This yields effective geometric part of the spherically symmetric Hamiltonian
\begin{equation}\label{eq:C_from_H_lift}
C[K, E] = \left(\frac{\partial_x \big( H[b, v, \mathcal{E}] \big)}{2 \sqrt{1+\mathcal{E}}} \right)\bigg|_{\mathcal{I}},
\end{equation}
which reproduces the spherically symmetric metric in generalised Painlev\'e-Gullstrand coordinates 
\begin{equation}\label{eq:gen_metric}
\mathrm{d}s^2 = -\mathrm{d} t^2 + \left( \frac{E^\varphi}{\sqrt{E^x}} \right)^2 (\mathrm{d}x + N^x \mathrm{d}t)^2 + E^x \mathrm{d}\Omega^2,
\end{equation}
where we have already implemented the partial temporal gauge fixing to comoving gauge. 

The structure of the constructive framework above offers a systematic way of constructing modified models beyond GR : we may generalise the shell Hamiltonian $H$ while maintaining the conservation of $\mathcal{E}$ and the form of the identification map. Any choice of polymerisation in the effective model is then encoded in the choice of the explicit function $H$.

We assume that the modification to the gravitational Hamiltonian dynamics is minimal, in the sense that it does not introduce any new propagating degrees of freedom other than the dust component itself.

Analogous to the mimetic gravity framework, we operate in the unitary gauge where the additional scalar degree of freedom is identified with the temporal coordinate of the comoving observer. This ensures that the total count of degrees of freedom remains 2 (tensor) + 1 (scalar/dust). Crucially, the scalar mode manifests solely as the pressureless dust fluid and does not propagate as a wave ($c_s = 0$).

Consequently, in a spherically symmetric configuration, where the tensor modes are suppressed. The absence of propagating degrees of freedom ensures the ultralocality of the gravitational dynamics. This structure guarantees that the total Hamiltonian can be decomposed into independent shell Hamiltonians one for each radial coordinate $x$, each parameterised by the conserved mass $M(x)$ and the LTB function $\mathcal{E}(x)$.

Modifying $H$ thus corresponds to modifying the effective gravitational dynamics sourced by the dust, while the geometric relationship between the LTB function $\mathcal{E}$ and triads and its derivatives respectively remains fixed, ensuring compatibility with the LTB ansatz. In the class of models we consider in this work the spatial diffemorphism on $x$ decouples from the dynamics and we assume it keeps its original form, thus we can safely add it to the Hamiltonian.

\section*{Assumption (i) Spatial Covariance: Resulting factorised Hamiltonian}
Next we restrict the possible effective models by means of the two principles (i) and (ii) and we start with (i), which will yield to restriction on the form of $H$ since the model must respect spatial diffeomorphism. In spherical symmetry, this implies that the Lagrangian density can depend only on scalar invariants constructed from the spatial geometry. The independent invariants are two build from the extrinsic curvature $\mathcal{K}_1$ and $\mathcal{K}_2$ with\footnote{All invariants can be expressed as polynomials of $\mathcal{K}_{1,2}$ and $\mathcal{R}_{1,2}$. For example, $K_i^i =\mathcal{K}_1 + \mathcal{K}_2, K_{ij} K^{ij} =3 (\mathcal{K}_1)^2 +  (\mathcal{K}_2)^2 - 2 \mathcal{K}_1 \mathcal{K}_2$. The three-dimensional Ricci scalar readas $R^{(3)} = 2\mathcal{R}_1 + 4\mathcal{R}_2$ and the Kretschmann scalar  has the form $\mathcal{K}^{(3)} = 4\mathcal{R}_1^2 + 8\mathcal{R}_2^2$.}
\begin{align}
  \mathcal{K}_1 \equiv K_{\theta}^{\theta} = \frac{\dot{E^{x}}}{2 {E^x}},\;\; \mathcal{K}_2 \equiv K_{x}^{x}+K_{\theta}^{\theta} = \frac{\dot{E^{x}}'}{{E^{x}}'}= \frac{\dot{E^{\phi}}}{{E^{\phi}}},
\end{align}
and two independent from the spatial curvature $\mathcal{R}_1, \mathcal{R}_2$ given by:
\begin{align}
    \mathcal{R}_1 \equiv - R^{\theta \phi}{}_{\theta \phi} =  \frac{\mathcal{E} }{E^x}, \quad
    \mathcal{R}_2 \equiv R^{x \theta}{}_{x \theta} =  -\frac{\mathcal{E}' }{{E^x}' }.
\end{align}
The invariant $\mathcal{R}_1$ is precisely the LTB function $\mathcal{E}$ scaled by the areal radius squared ($E^{x} \propto r^{2}$), representing a dimensionless measure of spatial curvature on the shell. This requirement severely restricts the form of the shell Hamilton  and ultimately forces it to adopt a \emph{factorised structure} of the following form:
\begin{align}\label{eq:H_separable}
H[b, v, \mathcal{E}] = v \, H_0\!\left(b, \mathcal{R}_1\right), \qquad \text{with} \quad \mathcal{R}_1 = \frac{\mathcal{E}}{v^{2/3}} .
\end{align}
It is noteworthy, that this automatically leads to the so-called $\bar{\mu}$ scheme in LQC.

The requirement that the Lagrangian depends exclusively on $\mathcal{K}_{1}, \mathcal{K}_{2}, \mathcal{R}_{1}, \mathcal{R}_{2}$ implies, after a detailed analysis of the Legendre transformation, that $b =\mathcal{B}(\mathcal{K}_1,\mathcal{R}_1 )$ can only depend on $\mathcal{K}_1$ and $\mathcal{R}_1$ and, consequently, $H$ must be factorised as in \eqref{eq:H_separable}. This factorisation reduces the functional dependence of the model to a two-dimensional function $H_0(b, \mathcal{R}_1)$, which significantly simplifies the problem. After a direct calculation, the fully \emph{spatially covariant effective Lagrangian density} reads \cite{Giesel:2026long}
\begin{align}\label{eq:final_Lagrangian}
\mathcal{L}^{(3)} &= \frac{\sqrt{h}}{2} \, \Bigg[ 2\mathcal{B} \left(\mathcal{K}_2 +\mathcal{K}_1 \right) - 3 H_0 + 2 (\mathcal{R}_1 + \mathcal{R}_2) \partial_{\mathcal{R}_1} H_0 \Bigg],%
\end{align}
where $h$ denotes the determinant of the spatial metric.
\section*{Assumption (ii) Geometric Guiding Principle: Resulting Unique Polymerised Vacuum Solution}
Beyond the spatial diffeomorphism symmetry, we impose (ii): the \emph{geometric guiding principle} that the vacuum solution of the theory should be static and uniquely determined by its mass. This reflects the expectation that a reasonable modified gravity theory should retain the defining properties of the classical Schwarzschild vacuum solution. Within the LTB description, a vacuum region comprises shells with constant mass $2 M \equiv H$ (and constant $\mathcal{E}$). A detailed analysis of the Ricci and Kretschmar as well as the Weyl scalar shows that in contrast to the classical case in GR, these quantities do depend on the LTB function. Demanding that these dependence is absent yields to (only) two independent partial differential equations for the shift vector. %
Solving this set of PDEs yields
\begin{align}\label{eq:shift_condition}
(N^{r})^{2} = \mathcal{E} + f(2M, r),
\end{align}
where $f$ depends only on the mass $M$ and the areal radius $r$. In the special case of GR where $f(2M, r)=\frac{2M}{r}$ \eqref{eq:shift_condition} agrees with the condition for the shift vector in \cite{Lasky:2006hq} and hence can be understood as the generalisation for models beyond GR. This condition guarantees that $\mathcal{E}$ cancels out in the spatial part of the metric when expressed in static coordinates, yielding the familiar form
\begin{align}\label{eq:static_metric}
ds^{2} = -\bigl(1-f(2M,r)\bigr)d\tau^{2} + \frac{dr^{2}}{1-f(2M,r)} + r^{2} d\Omega^{2} .
\end{align}
Physically, this principle demands that for the polymerised vacuum spacetime the coordinate invariant curvature scalars as well as the Weyl scalar are independent from the LTB ($\mathcal{E}(x)$), which can understood as a gauge artifact in the classical theory. In other words, we require the vacuum geometry of the modified theory is characterised solely by a constant mass $M$, mimicking the no-hair property in a broader context. Note that in contrast to the result in \cite{Cafaro:2024vrw}, where the vacuum solutions are parametrised by two parameters, here we can restore a generlised Birkhoff's theorem for this class of models.%

\section*{From Geometry to Dynamics: Linking the Hamiltonian to the Vacuum geometry}
The shift $N^{r}$ is kinematically linked to the shell motion via $N^{r} = -\dot{R}$. Combining the equation of motion $\dot{v} = \{v, H\} = 3/2 \partial_{b} H$ with $v = R^{3}$ gives $\dot{R} = (2R^2)^{-1} \partial_{b} H$ and therefore in PG coordinates leads to
\begin{align}\label{eq:N_from_dynamics}
(N^{r})^2 = \left((2r^2)^{-1} \partial_{b} H \right)^2 .
\end{align}
Now substituting the kinematic relation \eqref{eq:N_from_dynamics} into the geometric condition \eqref{eq:shift_condition} yields a PDE that couples the Hamiltonian to the vacuum geometry:
\begin{align}\label{eq:PDE_for_H}
\left(\partial_{b} H\right)^2 =  4 r^{4} \left( \mathcal{E} + f(H, r) \right) .
\end{align}
The general solution of this PDE introduces an integration function $c_1(r, \mathcal{E})$. However, this function parameterises a canonical gauge freedom (a specific redefinition of $b$) %
and leaves the physical equations of motion (modified Friedmann equation) unchanged. This freedom corresponds to a local affine shift in the variable $b$, which does not change the equations of motion and thus leaves the Lagrangian density unchanged except for a boundary term. We can therefore consistently set $c_1 \equiv 0$, which we do in the following.

\section*{Combining Principles (i) and (ii): The Universal Form of the Vacuum Solution}
We now combine the two principles (i) and (ii) by enforcing the spatially covariant form of the Hamiltonian in \eqref{eq:H_separable}. This allows us to obtain the universal form of the metric function of the vacuum solution already presented in \eqref{eq:main_result_boxed}. For this purpose we substitute $H = r^{3} H_0(b, \mathcal{R}_1)$ and $\mathcal{E} = r^{2} \mathcal{R}_1$ into \eqref{eq:PDE_for_H} leading to
\begin{align}\label{eq:combined}
\, \left(\partial_{b} H_0(b, \mathcal{R}_1) \right)^2 = 4 \left( \mathcal{R}_1 + \frac{f(r^{3} H_0, r)}{r^{2}} \right).
\end{align}
The left side of \eqref{eq:combined} explicitly depends only on the variables $b$ and $\mathcal{R}_1$. For reasons of consistency, the right side must therefore be expressed exclusively in terms of the same two variables. Since $\mathcal{R}_1$ already appears explicitly, the remaining term $f(r^{3} H_0, r)/r^{2}$ can only depend on $b$ via  $H_0(b, \mathcal{R}_1)$. Therefore, $f/r^{2}$ must be a function of $H_0$ only: $f(r^{3} H_0, r)/r^{2} = \tilde{f}(H_0)$. On a vacuum shell, the Hamilton is equal to the conserved mass, $r^{3} H_0 = 2 M$.
 We thus obtain the \emph{universal form for the static vacuum metric function} in our framework:
\begin{align}\label{eq:main_result}
\boxed{\; f(M, r) = r^{2} \; \widetilde{f}\!\left( \frac{2M}{r^{3}} \right) \; } .
\end{align}
The metric function $f$ is therefore $r^{2}$ times an arbitrary function $\widetilde{f}$ of the of the dimensionless mass density scale $2M/r^{3}$. 
General relativity and thus the Schwarzschild solution is recovered for the linear choice $\tilde{f}(X) = X$ with $X=\frac{2M}{r^3}$.

\section*{The Resulting Covariant Theory}
Our framework provides two main findings. First, \eqref{eq:main_result} provides a restrictive condition  on the form of static vacuum solutions in any modified gravity theory constructible via our LTB-based construction approach. Second, and equally important, the factorised Hamiltonian $H_0(b, \mathcal{R}_1)$ directly determines a completely \emph{spatially covariant effective action} for the model \eqref{eq:final_Lagrangian}. This action can be further extended to a \emph{four-dimensional covariant metric theory} by introducing a scalar field $\phi$, similar to ideas presented in \cite{Fujita:2015ymn,Gao:2020yzr}, see \cite{Giesel:2026long} for details. Note here the scalar field has a unit timelike gradient ($g^{\mu\nu}\partial_{\mu}\phi\partial_{\nu}\phi = -1$), which perserves the fact that the lapse function is $N=1$. This then follows the mimetic gravity paradigm, where extrinsic curvatures  can be represented as $\mathcal{P}(K_{ab})=\nabla_{a} \nabla_{b} \phi$ and three curvature invariants can be represented as combinations of four curvature invariants and covariant derivatives of $\phi$ \footnote{This is done by using Gau\ss{}-Codazzi equation and that in our gauge choice ($N=1$) the conormal to the hypersurface is given by $n_a = \nabla_a \phi$. For example, we have $
\mathcal{P}({R}^{(3)}) = {{R}^{(3)}} + 2 {{R}}^{(4)}_{ab} \nabla_a \phi \nabla_b \phi + \nabla_{a} \nabla_{b} \phi \nabla^{a} \nabla^{b} \phi - (\nabla^{a} \nabla_{a} \phi)^2$}. By substituting these relationships into the spatially covariant Lagrangian (\ref{eq:final_Lagrangian}), we obtain a four-dimensional covariant action of the form \cite{Giesel:2026long}
\begin{align}\label{eq:4d_Lagrangian}
\mathcal{L}^{(4)} &=\mathcal{P}(\mathcal{L}^{(3)}) + \frac{1}{2} \sqrt{|g|}\lambda \left(g^{\mu\nu}\partial_{\mu}\phi\partial_{\nu}\phi + 1 \right).
\end{align}
Interestingly, the resulting model \eqref{eq:4d_Lagrangian} belongs to a class of modified gravity theories beyond the standard extended mimetic models, as it includes couplings between the scalar field derivatives and the spacetime curvature. Since it originates from a spatially covariant action with second-order equations of motion, the covariant theory propagates the correct number of degrees of freedom (two tensorial plus one scalar) and avoids the Ostrogradsky instability.

This covariant action in \eqref{eq:4d_Lagrangian} opens a window to new insights for the application of perturbation theory around the unique vacuum backgrounds defined by \eqref{eq:main_result}, which consistently implement polymerisation effects in linear and higher orders. It finally provides a first-principles foundation needed to move beyond the hybrid approximation and compute, for example, quasinormal modes of polymerised black holes or the primordial power spectrum in a bouncing universe with non-zero spatial curvature, without ambiguity for the class of models that can be described by \eqref{eq:4d_Lagrangian}. 

\section*{Example: LQG Inspired Polymerised Models and Unique Curvature Coupling}
\subsection*{From flat-space polymerisation to a unique vacuum solution}
A common polymerisation ansatz in loop quantum cosmology, suitable for flat FLRW models ($\mathcal{E}=0$) \cite{Ashtekar:2023cod}, is
\begin{align}
    H_{\text{flat}} = \frac{v}{2G}\, \frac{\sin^2(\alpha b)}{\alpha^2},
\end{align}
where $\alpha$ is the polymerisation scale (of dimension length). In the classical limit $\alpha\to0$, this reduces to the Hamiltonian of GR $H_{\text{GR}}=(v/2G)\,b^{2}$.

When such a model is extended to non-marginally bound LTB spacetimes ($\mathcal{E}\neq0$), an ambiguity arises regarding the polymerisation of the curvature term $\mathcal{E}$. However, our constructive framework provides a clear and unambiguous answer. Suppose that the polymerised model allows for a unique static vacuum solution whose metric function is that given in the marginally bound case and known from previous works \cite{Kelly:2020uwj,Lewandowski:2022zce,Giesel:2023hys} to be
\begin{align}
    f(M,r) = \frac{2M}{r} - \frac{(2M)^{2}\alpha^{2}}{r^{4}} .
\end{align}
This can be written as $f(M,r)=r^{2}\tilde{f}\left(\frac{2M}{r^{3}}\right)$ with
\begin{align}
    \tilde{f}(X)=X\,(1-\alpha^{2}X),\quad X:=\frac{2M}{r^3}.
\end{align}
which satisfies the requirement (i) and (ii) and hence \eqref{eq:main_result}. This gives the polymerised Friedmann equation according to \eqref{eq:shift_condition}:
\begin{align}\label{eq:polymerized_Friedmann}
    \dot{R}^2 = (N^{r})^{2} = \mathcal{E} + r^{2}\tilde{f}\!\left(\frac{2M}{r^{3}}\right),
\end{align}
where we used that in PG coordinates we have $R=r$. Thus, the modified Friedmann equation is completely determined by (i) and (ii); there is no freedom left for different ways of polymerising the curvature terms.

It is worth noting that although this unique vacuum solution formally possesses a singularity at $r=0$, this feature is generated by the singular distributional source (point mass) located at the origin. However, as we will show, the effective dynamics ensure that any collapsing dust configuration with a non-trivial energy density undergoes a non-singular bounce. Thus, in any realistic physical collapse process, the singularity is resolved dynamically.

\subsection*{Reconstructing the Hamiltonian}
The entire power of the constructive framework can be also seen from the fact that it further determines the Hamiltonian that yields this vacuum solution. Substituting the factorised ansatz $H=v H_{0}(b,\mathcal{R}_{1})$ and the expression for $\tilde{f}$ into the fundamental relation \eqref{eq:combined} yields a first-order differential equation for $H_{0}$. Its general solution is given by
\begin{align}\label{eq:H0_polymerized}
    H_{0}(b,\mathcal{R}_{1}) = \frac{1}{2\alpha^{2}}\Bigl[1-
        \sqrt{1+4\alpha^{2}\mathcal{R}_{1}}\;
        \cos\!\bigl(2\alpha b + \alpha c_{1}(\mathcal{R}_{1})\bigr)\Bigr].
\end{align}
The function $c_{1}$ reflects the residual freedom in defining the variables $b$ on different shells we mentioned before.
In the case when $\mathcal{R}_1 \geq 0$, with a special choice of $c_{1}(\mathcal{R}_{1})$, the Hamiltonian can be rewritten in a form that makes the coupling to spatial curvature manifest:
\begin{align}
    H = \frac{v}{2G}\Bigg[ \frac{\sin^{2}\!\bigl(\alpha(b+\sqrt{\mathcal{R}_{1}})\bigr)}{\alpha^{2}}
        - \frac{\sqrt{\mathcal{R}_{1}}}{\alpha}\,
          \sin\!\bigl(2\alpha(b+\sqrt{\mathcal{R}_{1}})\bigr) \Bigg].
\end{align}
It is important to note that althrough this Hamiltonian only works for non-negative $\mathcal{R}_1$, the dynamics can be extended to negative $\mathcal{R}_1$ as well, as its generate the same modified Friedmann equation as \eqref{eq:polymerized_Friedmann}. For $\mathcal{R}_{1}=0$ (flat spatial sections) the second term vanishes and we recover the standard polymerised Hamiltonian $H_{\text{flat}}$. For non‑zero $\mathcal{R}_{1}$, the curvature enters in a minimal way: it simply shifts the argument of the polymerisation sine‑functions by $\sqrt{\mathcal{R}_{1}}$, plus an additional term that guarantees the factorised structure required by (i), that is spatial covariance.

Requiring the Hamiltonian and the metric components (specifically the shift vector) in PG coordinates to remain real-valued imposes the following consistency conditions:
\begin{align}
    \mathcal{R}_{1}+\tilde{f}(H_{0}) \geq 0, \;\; 1+4\alpha^{2}\mathcal{R}_{1} \geq 0 .
\end{align}
Dynamically, these conditions are satisfied as a direct consequence of the effective bounce, which restricts the physical evolution of the dust to the phase space region where these inequalities hold.

However, it is crucial to distinguish the validity of the coordinate chart from the validity of the spacetime solution. In the vacuum limit, these inequalities represent restrictions on the \emph{domain of validity of the comoving coordinate chart}, which is constructed from a specific family of test-particle geodesics parameterised by the energy function $\mathcal{E}$. Since these geodesics themselves may undergo an effective bounce, the coordinate chart covering them is naturally bounded. Indeed, different choices of $\mathcal{E}$ correspond to different domains of validity for the associated coordinate charts, as shown in Fig \ref{fig:plotev}. In the limit $\mathcal{E} \to \infty$, the chart effectively covers the entire spacetime region. This behavior underscores the fact that the static vacuum solution itself is unique and independent of the auxiliary parameter $\mathcal{E}$, which merely serves as a label for the choice of observer, in contrast to the models considered in \cite{Giesel:2024mps,Cafaro:2024vrw}. 

\subsection*{Resolution of the ambiguity}
The main result of this example is the Hamiltonian in \eqref{eq:H0_polymerized}. It provides the \emph{unique} polymerisation of the LTB shell Hamiltonian that complies with spatial diffeomorphism invariance and the geometric guiding principle for unique vacua. Any other prescription for coupling the curvature term $\mathcal{E}$ would either violate one of these principles (i) and (ii) or correspond to a different choice of the gauge function $c_{1}$, which merely redefines the variable $b$ without changing the physics. Consequently, the long‑standing ambiguity of how to polymerise curvature in LQG inspired models \cite{Corichi:2011pg,Singh:2013ava,Langlois:2017hdf} is removed: the consistent extension from flat to curved spatial slices is dictated by the fundamental principles (i) and (ii) embodied in our constructive framework.

The Hamiltonian~\eqref{eq:H0_polymerized} leads, via the identification map, to a spatially covariant action, invariant under spatial diffeomorphisms, that can be extended to a full four‑dimensional metric theory with a scalar field (generalised mimetic‑type). These kind of covariant models are free of Ostrogradsky instabilities and provides the necessary foundation for consistent perturbation theory investigations around the unique polymerised black‑hole or cosmological backgrounds which includes polymerisation effects also in linear and higher orders. 

\subsection*{Properties of the bounce for models with non-zero spatial curvature}
\begin{figure}[h]
    \includegraphics[width=0.8\columnwidth]{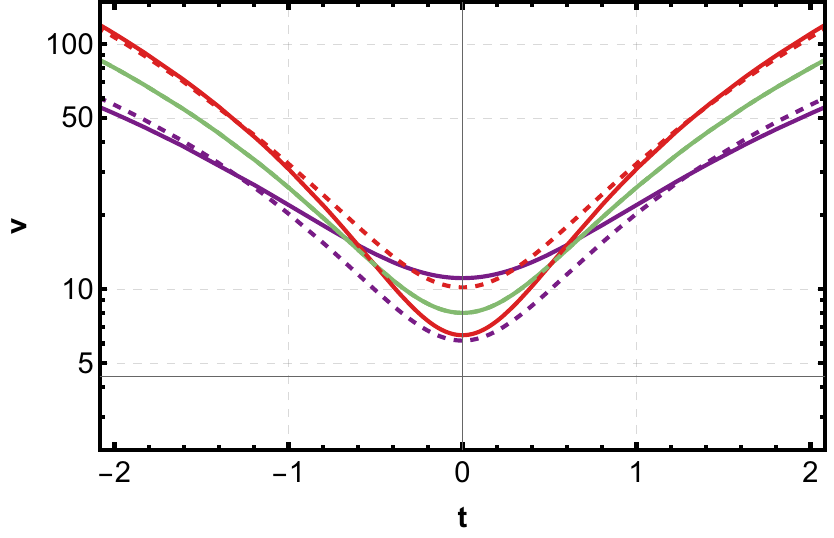}
    \caption{This plot compares different bouncing dynamics for dynamics from this work (thick line) and standard curvature polymerization in LQC (dashed line) with $\mathcal{E}=-1$ (purple lines), $\mathcal{E}=0$ (green lines) and $\mathcal{E}=1$ (red lines). The dynamics show similar behaviour far away from the bounce where classical physics is dominating, while the dynamics changes for the deep quantum regime in the non-marginally bound case. }
    \label{fig:plotvt}
\end{figure}

\begin{figure}[h]
    \includegraphics[width=0.8\columnwidth]{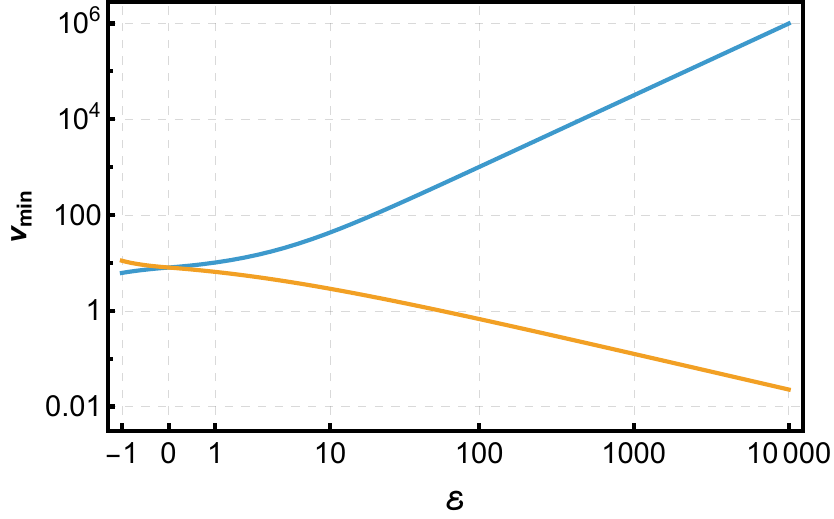}
    \caption{This plot shows that how the minimal volume changes with different LTB function $\mathcal{E}$ for a fixed mass $m = 4$ and $\alpha=1$. The new dynamics (yellow line) shows a different behavior compare to the standard curvature polymerization in LQC dynamics. With a decreasing bouncing radius for increasing $\mathcal{E}$, the bounce disappear in the limit $\mathcal{E} \to \infty$. }
     \label{fig:plotev}
\end{figure}

If we consider again the more general case where the dust energy density is non-vanishing, the effective dynamics generated by the Hamiltonian~\eqref{eq:H0_polymerized} leads to a non-singular bounce for collapsing dust shells. The bounce occurs when $\dot{R}=0$, which from \eqref{eq:polymerized_Friedmann} implies
\begin{align}\label{eq:bounce_condition}
    \mathcal{E} + r^{2}\tilde{f}\!\left(\frac{2M}{r^{3}}\right) = 0.
\end{align}
Note that $\dot{R}^2$ is always non-negative. This condition sets a lower bound on possible $\mathcal{E} \geq \text{max}\left\{-1,-\frac{3 M^{2/3}}{4 \alpha ^{2/3}}\right\}$. Under this condition, the 4th order equation \eqref{eq:bounce_condition} always admits two possibly degenerate real roots. Actually according to Descartes' Rule of Signs, in case $\mathcal{E} \geq 0$, we always have one positive roots while for $\mathcal{E} < 0$ we have two positive roots. The shell bounces at the smaller positive root $R_{B}$, which is always finite and non-zero for non-zero $\alpha$, thus resolving the classical singularity at $R=0$ with a symmetric bounce. The larger root for $\mathcal{E} < 0$ represents the recollapse of the closed universe. The presence of spatial curvature $\mathcal{E}$ modifies the bounce radius compared to the marginally bound case, but does not eliminate the bounce itself, provided the lower bound on $\mathcal{E}$ is respected. In contrary to standard extrinsic curvature polymerisation in LQC \cite{Corichi:2011pg,Singh:2013ava,Giesel:2022rxi}, this new approaches here predicts a decreasing bouncing radius with increasing $\mathcal{E}$, as shown in Fig. \ref{fig:plotvt} and Fig. \ref{fig:plotev}. Note that for the model investigated in \cite{Giesel:2021dug} the standard LQC curvature polymerisation was used, which is not consistent with the two principles (i) and (ii). However, since that model contains polymerisations of both extrinsic curvature variables it does not belong to the class considered here.

\section*{Discussion and Conclusions}
In this work we introduced a constructive framework which allows to embed effective models into $4D$ covariant generalised extended mimetic gravity models, which have a unique vacuum solution, establishing a Birkhoff-like theorem for these class of models. By this we provide a solution to the long-standing presence of the ambiguity for the vacuum solution in the context of effective spherical dust-collapse models inspired by quantum gravity. By imposing spatial (i) diffeomorphism invariance and (ii) a geometric guiding principle demanding static, unique vacuum solutions, the shell Hamiltonian is forced into a factorised form 
$H = v H_0(b.\mathcal{R}_1)$, and the static vacuum metric is universally fixed as $f(2M,r) = r^2 \tilde{f}(\frac{2M}{r^3})$, where 
$\tilde{f}(X)=X$ recovers the Schwarzschild solution, and non-linear $\tilde{f}$ encode quantum gravity inspired corrections. This universal form provides a first-principles guiding criterion for modified gravity theories in this class and allows to embed those models via an extension into a ghost-free fully covariant four-dimensional action. This covariant four-dimensional action provides a consistent basis for perturbation theory with given cosmological or spherically symmetric backgrounds.

As a concrete illustration, the constructive framework uniquely determines how LQG-inspired polymerisations incorporate spatial curvature, thereby eliminating the ambiguity regarding the polymerisation of curvature terms.

Our results provide an interesting starting point for future phenomenological investigations: The metric function 
$f(M,r)$ offers a direct target for phenomenological tests via black-hole shadows, quasinormal modes and cosmological perturbations. Future work will include to extend the framework to other matter than dust, e.g. with standard scalar field thus extend the theory beyond the mimetic framework to more general scalar-tensor theory, compute the linear-perturbation spectrum of polymerised black holes, and derive the primordial power spectrum in bouncing cosmologies with spatial curvature.

\bibliographystyle{unsrtnat}
\bibliography{refs}

\end{document}

%% file: defs.tex
 \newcommand{\bq}{\begin{equation}}
 \newcommand{\eq}{\end{equation}}
 \newcommand{\bqn}{\begin{eqnarray}}
 \newcommand{\eqn}{\end{eqnarray}}

\NewDocumentCommand{\evalat}{sO{\big}mm}{%
  \IfBooleanTF{#1}
   {\mleft. #3 \mright|_{#4}}
   {#3#2|_{#4}}%
}

\def\be{\begin{eqnarray}}
\def\ee{\end{eqnarray}}

\newcommand{\sgn}{\mathrm{sgn}}